\documentclass[10pt,aps]{revtex4}

\usepackage{amsmath,amssymb,amsfonts,dcolumn,color,graphicx,graphics,latexsym,placeins,epsfig}
\newcommand{\be}{\begin{eqnarray}}
\newcommand{\ee}{\end{eqnarray}}
\newcommand{\ba}{\begin{eqnarray}}
\newcommand{\ea}{\end{eqnarray}}

\begin{document}

\title{\bf Entanglement degradation in multi-event horizon spacetimes}
\author{Sourav Bhattacharya}
\email{sbhatta@iitrpr.ac.in}
\author{Nitin Joshi}
\email{2018phz0014@iitrpr.ac.in}
\affiliation{\small  Department of Physics, Indian Institute of Technology Ropar,  Rupnagar, Punjab 140 001,
India\\}

\date{\today}

\begin{abstract}
\noindent
We investigate the degradation of quantum entanglement in the Schwarzschild-de Sitter black hole spacetime, by studying the mutual information and the logarithmic negativity for maximally entangled, bipartite  states for  massless minimal scalar fields. This spacetime is endowed with  a black hole as well as a cosmological event horizon, giving rise to particle creation at two different temperatures.  We consider two independent descriptions of thermodynamics and particle creation in this background. The first involves thermal equilibrium of an observer with  either of the horizons. We show that as of the asymptotically flat/anti-de Sitter black holes, in this case the entanglement or correlation degrades  with increasing Hawking temperatures. The second treats both the horizons combined in order to define a total entropy and an effective equilibrium temperature. We present a field theoretic derivation of this effective temperature  and show that unlike the usual cases, the particle creation does not occur here  in causally disconnected spacetime wedges but instead in a single  region. Using the associated vacua, we show that in this scenario the entanglement never degrades but increases with increasing  black hole temperature and holds true no matter how hot the black hole becomes or how small the cosmological constant is.  We argue that this  phenomenon can have no analogue in the asymptotically flat/anti-de Sitter black hole spacetimes.     \\
\end{abstract}
\maketitle

\maketitle
\section{Introduction}\label{S1}
\noindent
The study of quantum  entanglement between created particle pairs in relativistic backgrounds has received considerable attention in recent years. Such investigations chiefly involve the Rindler or non-extremal black holes, cosmological spacetimes and even the Schwinger pair creation mechanism, e.g.~\cite{mann}-\cite{Bhattacharya:2020sjr} and references therein. In the Rindler or  a non-extremal black hole background, particle creation occurs in causally disconnected spacetime wedges. Since the created particles are thermal, the associated randomness destroys the entanglement or quantum correlation between entangled states   as the black hole evaporates and gets hotter, first shown in~\cite{mann} and subsequently in e.g.~\cite{Pan:2008yi}-\cite{Mirza}.  

To the best of our knowledge, all such earlier studies were made in  asymptotically flat spacetimes. However, keeping in mind the observed accelerated expansion of our current universe, it is physically important to ask : how does such degradation get affected in the presence of a positive cosmological constant, $\Lambda$? Such spacetimes  can also model primordial black holes formed in the early inflationary universe, e.g.~\cite{Gibbons}. The chief qualitative difference of these  black holes with that of $\Lambda\leq 0$ is the existence of the cosmological event horizon for the former, an additional event horizon serving as the outer causal boundary of our universe. These two-event horizon spacetimes admit 
two-temperature thermodynamics and hence are qualitatively much different compared to the single horizon $\Lambda \leq 0$ cases, e.g.~\cite{Gibbons}-\cite{pappas}. With this motivation,  we wish to investigate in this paper the role of this two temperature particle creation in  the entanglement degradation.  Our chief goal here is to see whether in this physically well motivated spacetime, the multi-horizon structure brings in any {\it qualitatively new} feature compared to that of the single horizon, i.e. the $\Lambda \leq 0$ cases.

In the next section we outline very briefly the causal structure of the Schwarzschild-de Sitter spacetime, a static and spherically symmetric black hole  located in the de Sitter universe. In Sec.~3, we discuss the entanglement degradation in the thermodynamical setup proposed in~\cite{Gibbons},  where an observer can be in thermal equilibrium with either of the horizons and show that the results qualitatively resemble with that of the single horizon  spacetimes. We use  the mutual information and logarithmic negativity for a maximally entangled, bipartite Kruskal-like state corresponding to massless minimal scalar fields as appropriate measures. In Sec.~4, we adopt the so called total entropy-effective equilibrium temperature description to treat both the horizons combined, e.g.~\cite{Maeda}-\cite{Bhattacharya:2015mja}. We first present a field theoretic derivation of the effective temperature and show that  unlike the previous cases, the entangled pair creation in this scenario {\it does not} occur in causally disconnected wedges in the extended spacetime. Most importantly, we demonstrate that   the entanglement here actually {\it increases} with the increase in the black hole Hawking temperature, no matter how hot the black hole becomes or how small the cosmological constant is. We emphasise that this phenomenon  is purely an outcome of the two-horizon geometry and hence has no $\Lambda \leq 0$ analogue. We shall set $c=\hbar=k_{\rm B}=G=1$ throughout below.

\section{The basic setup}\label{S2}
\noindent
The Schwarzschild-de Sitter (SdS) spacetime, 
\begin{eqnarray}
ds^2=-\left(1-\frac{2M}{r}-\frac{\Lambda r^2}{3}\right)dt^2+\left(1-\frac{2M}{r}-\frac{\Lambda r^2}{3}\right)^{-1}dr^2+r^2 \left(d\theta^2 +\sin^2\theta d\phi^2 \right)
\label{l1}
\end{eqnarray}
admits  three event or Killing horizons for $0<3M \sqrt{\Lambda} < 1$, e.g.~\cite{Gibbons, JHT}, 
\begin{eqnarray}
r_{H}=\frac{2}{{\sqrt \Lambda} }\cos\frac{\pi+\cos^{-1}(3M\sqrt{\Lambda})}{3},~
r_{C}=\frac{2}{{\sqrt \Lambda} }\cos\frac{\pi-\cos^{-1}(3M \sqrt{\Lambda})}{3},~r_{U}=-(r_H+r_C)
\label{l2}
\end{eqnarray}
$r_H < r_C$ are respectively the black hole and the cosmological event horizon (BEH and CEH), whereas $r_{U}<0$ is unphysical. As $3M\sqrt{\Lambda} \to 1$ we have $r_H \to r_C$, known as the Nariai limit whereas for $3M\sqrt{\Lambda}>1$,  the spacetime is naked singular. Thus unlike $\Lambda \leq 0$, a black hole cannot  be arbitrarily massive here, for a given $\Lambda$. 

The surface gravities of BEH and CEH are respectively given by,
\begin{eqnarray}
\kappa_H=    \frac{\Lambda (2r_H+r_C)(r_C-r_H)}{6 r_H}, \quad 
-\kappa_C=\frac{\Lambda (2r_C+r_H)(r_H-r_C)}{6 r_C}
\label{l3}
\end{eqnarray}
Due to the repulsive effects generated by a  positive $\Lambda$, the surface gravity of CEH is negative.

Since  $r=r_H,\,r_C$ are two coordinate singularities, we need two Kruskal-like coordinates in order to  extend the spacetime beyond them, 
\begin{eqnarray}
ds^2=-\frac{2M}{r}\left\vert1-\frac{r}{r_C}\right\vert^{1+\frac{\kappa_H}{\kappa_C}} \left(1+\frac{r}{r_H+r_C}\right)^{1-\frac{\kappa_H}{\kappa_U}}\, d{\overline u}_H d {\overline v}_H+r^2(d\theta^2+\sin^2\theta d\phi^2)
\label{ds16}
\end{eqnarray}
and
\begin{eqnarray}
ds^2=-\frac{2M}{r}\left\vert\frac{r}{r_H}-1\right\vert^{1+\frac{\kappa_C}{\kappa_H}} \left(1+\frac{r}{r_H+r_C}\right)^{1+\frac{\kappa_C}{\kappa_U}}\, d {\overline u}_C d {\overline v}_C+r^2(d\theta^2+\sin^2\theta d\phi^2)
\label{ds17}
\end{eqnarray}
where,
\begin{eqnarray}
{\overline u}_H=-\frac{1}{\kappa_H}e^{-\kappa_H u},\quad {\overline v}_H=\frac{1}{\kappa_H}e^{\kappa_H v} \quad {\rm and} \quad
{\overline u}_C=\frac{1}{\kappa_C}e^{\kappa_C u},\quad {\overline v}_C=-\frac{1}{\kappa_C}e^{-\kappa_C v}
\label{ds15}
\end{eqnarray}
are the Kruskal null coordinates whereas $u=t-r_{\star}$ and $v=t+r_{\star}$ are the usual retarded and advanced null coordinates. The radial tortoise coordinate $r_{\star}$ is given by
\begin{eqnarray}
r_{\star}=\frac{1}{2\kappa_H}\ln \left\vert\frac{r}{r_H}-1\right\vert -\frac{1}{2\kappa_C} \ln \left\vert1-\frac{r}{r_C}\right\vert +\frac{1}{2\kappa_U}\ln \left\vert\frac{r}{r_U}-1\right\vert
\label{ds5}
\end{eqnarray}
$\kappa_U$ is the `surface gravity' of the unphysical horizon located at $r_U =-(r_H+r_C)$. Note that Eq.~(\ref{ds16}) and Eq.~(\ref{ds17}) are  free of coordinate singularities respectively on the BEH and CEH.  However, there is no single Kruskal coordinate for the SdS spacetime that simultaneously removes the coordinate singularities of both the horizons. 

\begin{figure}[h!]
\centering
  \includegraphics[width=9cm]{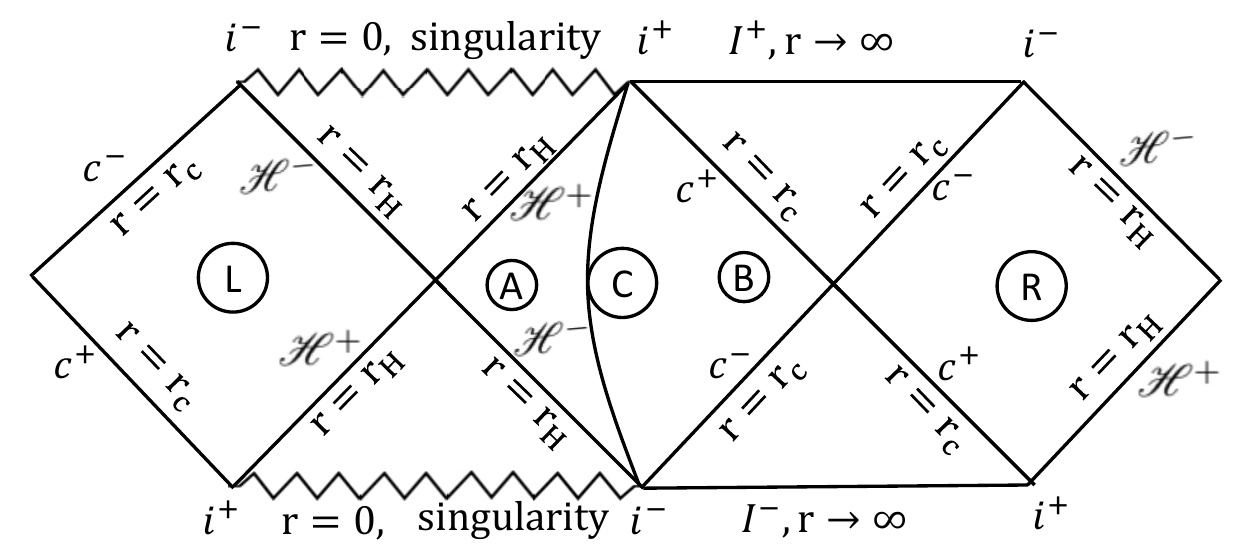}
  \caption{\footnotesize The Penrose-Carter diagram of the  extended Schwarzschild-de Sitter spacetime. ${\cal H }^{\pm}({\cal C}^{\pm})$ respectively denote the future and past black hole event horizons (cosmological event horizons). 
$i^{\pm}$ respectively represent the future and past timelike infinities, whereas the infinities $ I^{\pm}$ are spacelike.
    The regions R, L are time reversed with respect to C and all the seven wedges are causally disconnected. The spacetime can further be extended towards both sides indefinitely, but we do not require this for our current purpose. Our region of interest is C ($r_H< r < r_C$) and hence we shall trace over the states belonging to other regions when it is  relevant. The hyperbola joining $i^{\pm}$ is a thermally opaque membrane separating the region C into two subregions A and B (${\rm C}={\rm A}\cup {\rm B}$), as discussed in Sec.~\ref{S3}. The hyperbola is drawn with respect to the black hole Kruskal coordinate (the first of Eqs.~(\ref{ds5'})), but can be drawn with respect to cosmological Kruskal coordinates as well. In Sec.~\ref{s4} however, such membrane will not be considered.}
  \label{fl1}
\end{figure}

Finally, by defining the Kruskal timelike and spacelike coordinates as,
$$\overline{u}_H = T_H -R_H,\quad \overline{v}_H = T_H+R_H,\qquad {\rm and} \qquad \overline{u}_C = T_C -R_C,\quad \overline{v}_C = T_C+R_C, $$
and using Eq.~(\ref{ds15}), we respectively have the relations
\begin{eqnarray}
&&-\overline{u}_H \overline{v}_H=R_H^2 -T_H^2= \frac{1}{\kappa_H^2} \left\vert 1-\frac{r}{r_C}\right\vert^{-\kappa_H/\kappa_C} \left\vert \frac{r}{r_U}-1\right\vert^{\kappa_H/\kappa_C}\left(\frac{r}{r_H}-1 \right)\nonumber\\
&&-\overline{u}_C \overline{v}_C=R_C^2 -T_C^2= -\frac{1}{\kappa_C^2} \left\vert \frac{r}{r_U}-1\right\vert^{-\kappa_C/\kappa_U} \left\vert \frac{r}{r_H}-1\right\vert^{-\kappa_C/\kappa_H}\left(1-\frac{r}{r_C} \right)
\label{ds5'}
\end{eqnarray}
Thus with respect to either of the above Kruskal coordinates, an $r={\rm const.}$ line is a hyperbola. Fig.~(\ref{fl1}) shows the Penrose-Carter diagram of the maximally extended SdS spacetime. 
%

\section{The Gibbons-Hawking thermal states and entanglement} \label{S3}
\noindent
The two event horizons of the SdS spacetime produce two thermodynamic relationships with temperatures $\kappa_H/2\pi$ and $\kappa_C/2\pi$,
\be
\delta M = \frac{\kappa_H}{2\pi} \frac{\delta A_H}{4}, \qquad \delta M= -\frac{\kappa_C}{2\pi} \frac{\delta A_C}{4} 
\label{l4}
\ee
where $A_H$ and $A_C$ are respectively the areas of the BEH  and CEH. However, since $r_C\geq r_H$, we have  $\kappa_H \geq \kappa_C$, Eq.~(\ref{l3}), and accordingly one expects that any equilibrium  is not possible. 
One way to tackle this issue is to place a thermally opaque membrane in the region C in Fig.~(\ref{fl1}), thereby splitting it into two thermally isolated subregions~\cite{Gibbons}.  Thus an observer located at the black hole side detects  Hawking radiation at temperature $\kappa_H/2\pi$ and likewise another on the other side detects   the same at temperature $\kappa_C/2\pi$. Note in analogy that even in an asymptotically flat black hole spacetime, in order to define the Hartle-Hawking state, which describes thermal equilibrium of the black hole with a blackbody radiation at its Hawking temperature, one also needs  to `encase' the black hole with a perfectly thermally reflecting membrane~\cite{BD}.

The existence of such states in the SdS spacetime was explicitly demonstrated in~\cite{Gibbons} via the path integral quantisation.  For our purpose, we shall very briefly demonstrate below their existence via the canonical quantisation. For the sake of simplicity, we shall work below in $(1+1)$-dimensions,  and consider a  free, massless and minimally coupled  scalar field,
 $\Box \phi(x)=0$. In any of the coordinate systems, the mode functions are simply plane waves.

Let us first consider the side of the  membrane which faces the BEH and call this subregion as A. The field quantisation can be done in a manner similar to that of the Unruh effect~\cite{Unruh:1976db, Crispino:2007eb}, and we shall not go into detail of it here. The local modes correspond to the $t-r_{\star}$ coordinates in A and also in the causally disconnected region L (with the time $t$ reversed) in Fig.~(\ref{fl1}). Note that there are both right and left moving plane wave modes characterised by the retarded and advanced null coordinates $u$ and $v$. The field quantisation can be done with both these kind of positive and negative frequency modes. However, since the left and right moving modes are orthogonal, the creation and annihilation operators associated with these two sectors commute. Accordingly, these two sectors can be treated as independent and without any loss of generality, we may focus on only one sector, e.g.~\cite{Crispino:2007eb}. This field quantisation yields the local vacuum, $|0_{A}, 0_{L}\rangle $. The global vacuum, $|0\rangle_{\kappa_H}$, in A$\cup$L corresponds to the field quantisation with the Kruskal coordinate of Eq.~(\ref{ds16}), regular on or across the BEH. Likewise, by calling the other subregion of C as B, we use the $t-r_{\star}$ coordinate and  Eq.~(\ref{ds17}) to make the field quantisation in B$\cup$R.  We have the Bogoliubov relationships and accordingly the squeezed state expansion similar to that of the Rindler spacetime, 
\be
|0\rangle_{\kappa_H}=\sum_{n=0}^{\infty} \frac{\tanh^n {r}}{\cosh{r}}|n_{A}, n_{L}\rangle \quad {\rm and }\quad  |0\rangle_{\kappa_C}=\sum_{n=0}^{\infty} \frac{\tanh^n {s}}{\cosh{s}}|n_{B}, n_{R}\rangle
\label{l5}
\ee
where $\tanh{r}=e^{-\pi \omega/\kappa_{H}}$ and $\tanh s= e^{-\pi \omega/\kappa_{C}}$. In other words, the Kruskal or the global vacuum states are analogous to that of the Minkowski vacuum, whereas the states appearing on the right hand side of the above equations are analogous to that of the local Rindler states confined to some particular spacetime regions. 

The squeezed state expansions of Eq.~(\ref{l5}) correspond  to  Planck spectra of created particle pairs respectively with temperatures $\kappa_H/2\pi$ and $\kappa_C/2\pi$  in regions A$\cup$L and B$\cup$R. These spectra are detectable respectively by observers located at the black hole and the cosmological horizon side of the thermally opaque membrane.  As we mentioned earlier, this setup was first proposed in~\cite{Gibbons} and accordingly, we shall regard these states as the Gibbons-Hawking thermal states. 

 For our purpose of forming entangled states, we shall also require the one particle  excitations $|1\rangle_{\kappa_H}$ and $|1\rangle_{\kappa_C}$, found by applying once the relevant creation operator on $|0\rangle_{\kappa_H}$ and $|0\rangle_{\kappa_C}$.  Using Eq.~(\ref{l5}) and the Bogoliubov relations we re-express these one particle states in terms of the local squeezed states,
\begin{eqnarray}
|1\rangle_{\kappa_H}=\sum_{n=0}^{\infty} \frac{\tanh^n {r}}{\cosh^2{r}} \sqrt{n+1}  |(n+1)_{A}, n_L\rangle,\qquad
|1\rangle_{\kappa_C}=\sum_{n=0}^{\infty} \frac{\tanh^n {s}}{\cosh^2{s}} \sqrt{n+1}|(n+1)_{B},n_R \rangle
\label{onep}
\end{eqnarray}

Note that even though the  construction of the Gibbons-Hawking states is mathematically consistent, we may wonder how one may {\it practically} realise such a thermally opaque membrane between the two horizons. Perhaps one possible way to realise this will be to consider the Klein-Gordon equation in $3+1$-dimensions, with the radial function satisfying,
$$ \left(-\frac{\partial^2}{\partial t^2}+\frac{\partial^2}{\partial r_{\star}^2} \right)R(r)+\left(1-\frac{2M}{r}-\frac{\Lambda r^2}{3} \right)\left(\frac{l(l+1)}{r^2} +\frac{2M}{r^3} -\frac{\Lambda}{3} \right) R(r)=0$$
The effective potential term appearing in the above Schr\"{o}dinger-like equation  vanishes at both the horizons and is positive in between. This bell shaped potential thus will work as a barrier between the two horizons. Modes that cannot penetrate it, will be confined in the regions close to the horizons and hence will be disconnected from each other. The effective potential thus can be thought of as a natural realisation of the thermally opaque membrane mentioned above.

Let us now take a maximally entangled global state,
\begin{eqnarray}
|\psi\rangle=\frac{1}{\sqrt{2}}\left[|0_{\kappa_H}, 0_{\kappa_C}\rangle+|1_{\kappa_H},  1_{\kappa_C}\rangle\right],
\label{x1}
\end{eqnarray}
and imagine that the $\kappa_H$- and $\kappa_C$-type states are located respectively in subregions A and B of C, defined above. Due to this placement, we can use  Eqs.~(\ref{l5}),~(\ref{onep}) to consistently re-express $|\psi \rangle$ in terms of the local states in regions L, A and  B, R. Tracing out now the states belonging to the  causally disconnected regions R and L, Fig.~(\ref{fl1}),   the reduced density operator for $|\psi \rangle $ becomes,
\begin{eqnarray}
\rho_{AB}&=\frac{1}{{2}} \sum_{n,m=0}^{\infty} \frac{\tanh^{2n} {r}\tanh^{2m} {s}}{\cosh^2{r} \cosh^2{s}}  \left[|n,m\rangle\langle n,m|+\frac{\sqrt{(n+1)(m+1)}}{{{\cosh{r}}\,{\cosh{s}}}}|n,m\rangle\langle n+1, m+1|\right.\nonumber\\&\left.+\frac{\sqrt{(n+1)(m+1)}}{{{\cosh{r}}\,{\cosh{s}}}} |n+1, m+1\rangle\langle n,m| +\frac{{(n+1)}{(m+1)}}{{{\cosh^2{r}}{\cosh^2{s}}}} |n+1, m+1\rangle\langle n+1, m+1|\right]
\label{x2}
\end{eqnarray}
where in any ket or bra, the first and second entries respectively belong to  A and B. With the help of this reduced, bipartite and mixed density matrix, we shall compute two appropriate measures of quantum entanglement  -- the mutual information and the logarithmic negativity (e.g.~\cite{NC} and references therein for detail). 

\begin{figure}[h!]
\begin{center}
  \includegraphics[width=6.5cm]{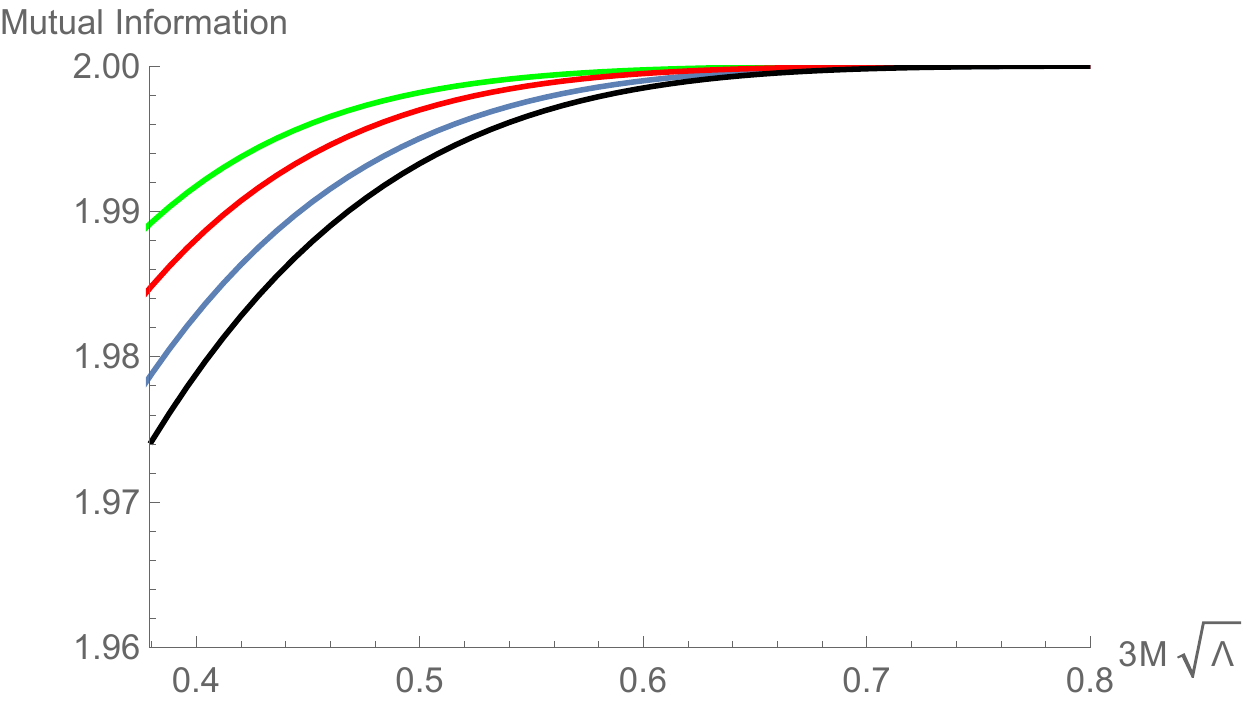}
  \includegraphics[width=6.5cm]{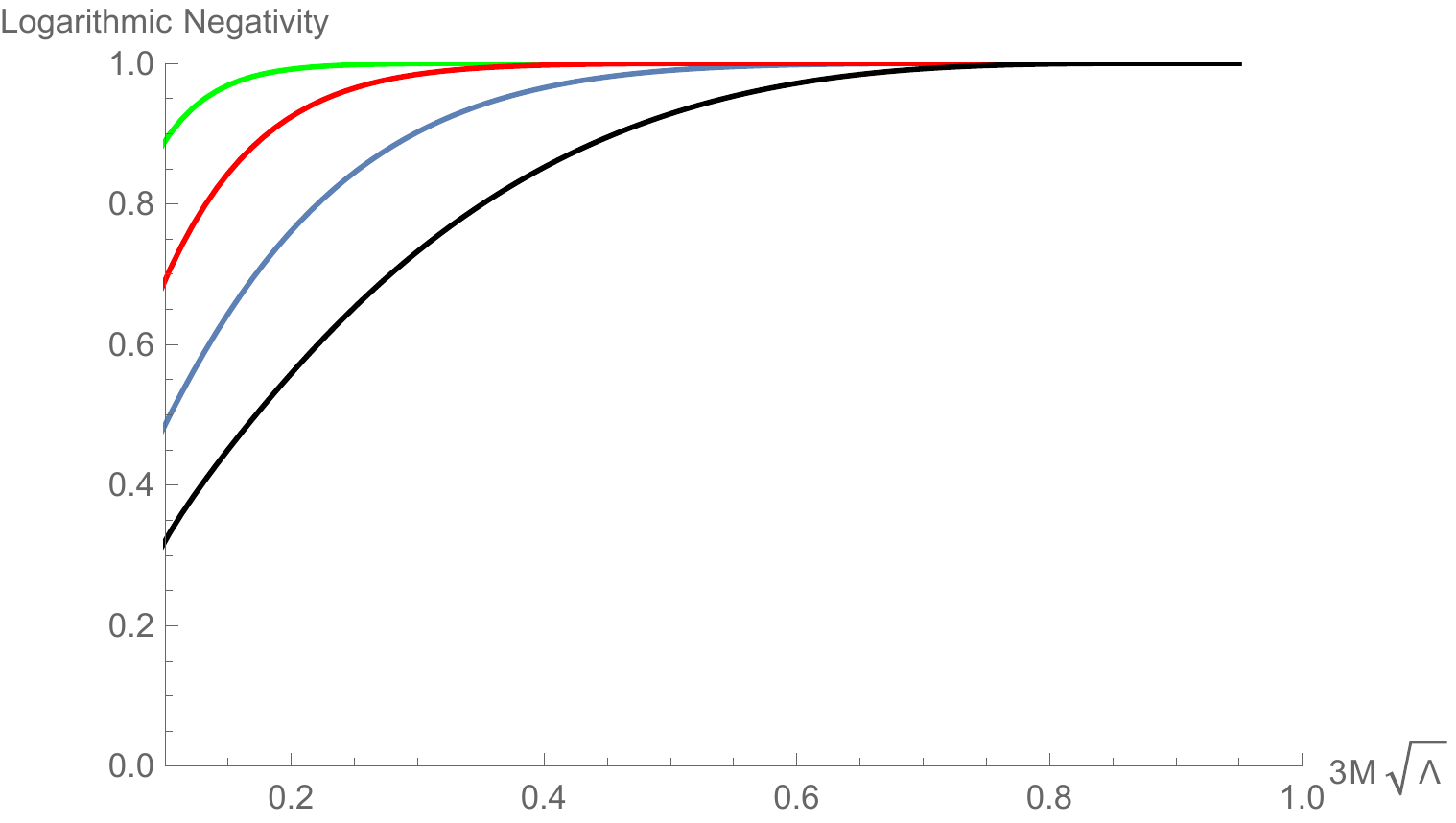}
  \caption{(Left) The mutual information for the maximally entangled state
 ${|\psi \rangle }$, Eq.~(\ref{x1}), vs. the dimensionless parameter $3M\sqrt{\Lambda}$.  The black,  blue, red, green curves correspond respectively to $\omega/ \sqrt{\Lambda}  = 0.94, 1,\,1.1,\,1.2$. (Right) The logarithmic negativity vs. $3M\sqrt{\Lambda}$. The black. blue, red, green curves correspond respectively to $\omega/ \sqrt{\Lambda}  = 0.6, 1,\,1.7,\,3$. For smaller values of $\omega/ \sqrt{\Lambda}$, the  curves will show further smaller values of the information quantities, for a given $3M\sqrt{\Lambda}$. This is because such lesser values (for a fixed $\omega$) would correspond to higher $\Lambda$ values, leading to higher rate of cosmological particle creation, eventually degrading the entanglement more. See main text for discussion.  }

  \label{fig2}
\end{center}
\end{figure}

The quantum mutual information of A and B is defined as
\begin{eqnarray}
{\cal{I}}_{AB}=S(\rho_{A})+S(\rho_{B})-S(\rho_{AB}),
\label{c1}
\end{eqnarray}
where $S= - {\rm Tr} (\rho \ln \rho)$ is the von Neuman entropy. Tracing out further the states belonging to  the subregion B, and A, we respectively have, 
\begin{eqnarray}
\rho_{A}
=\frac{1}{{2}} \sum_{n=0}^{\infty} \frac{\tanh^{2n} {r}   }{\cosh^2{r}  } \left(1+\frac{n}{\sinh^2{r}}\right) |n\rangle\langle n|\quad {\rm and} \quad
\rho_{B}
=\frac{1}{{2}} \sum_{m=0}^{\infty} \frac{\tanh^{2m} s}{\cosh^2s}\left(1+\frac{m}{\sinh^2{s}}\right) |m\rangle\langle m|
\label{c3}
\end{eqnarray}
Using Eq.~(\ref{x2}), Eqs.~(\ref{c3}), we now compute ${\cal{I}}_{AB}$ numerically in Mathematica.  Implicitly assuming $\Lambda$ to be fixed, the variation of ${\cal{I}}_{AB}$ with respect to the dimensionless parameter $3M{\sqrt \Lambda}$  is depicted   in the first of Figs.~(\ref{fig2}) for three different values of the dimensionless parameter $\omega/\sqrt{\Lambda}$. Thus the mutual information increases monotonically with increasing $M$ and  saturates to two in the Nariai limit,  $3M\sqrt{\Lambda}\to 1$. Eq.~(\ref{l3}), Eq.~(\ref{l2}) shows that the increase in this parameter corresponds to  the decrease in both the surface gravities, eventually becoming vanishing in the Nariai limit.
On the other hand as $3M\sqrt{\Lambda}\to 0$, we have $\kappa_H \sim M^{-1}$ and $\kappa_C \sim \sqrt{\Lambda}$. Thus for a fixed value of $\Lambda$,  black hole's Hawking  temperature becomes very large in this limit, resulting in an extremely high rate of particle creation.

Intuitively, the increase in the Hawking temperature increases the degree of randomness of the created thermal particles, which  degrades or destroys quantum correlation, as has been reflected in Figs.~(\ref{fig2}), analogous to that of the single horizon spacetimes reported earlier~\cite{mann}-\cite{Mirza}. We also note that for a given value of $\omega$, the mutual information   degrades more with increasing $\Lambda$. This  is due to the increase in the   Hawking temperature of the CEH  with increasing $\Lambda$,  resulting in  degradation  in the  correlation further, compared to the $\Lambda=0$ case.  

The logarithmic negativity  is related to the eigenvalues of Eq.~(\ref{x2}), but after transposing one of sectors (say A), $(\rho_{AB})^{T_{A}}$. It is defined as ${\cal{L_{N}}}=\log\sum_i |\lambda_{i}|$,  where $\lambda_{i}$'s are the eigenvalues $(\rho_{AB})^{T_{A}}$, e.g.~\cite{NC}. 
We have also computed it numerically in Mathematica and have plotted in the second of Figs.~(\ref{fig2}). The qualitative conclusions remain the same as that of the mutual information. 

 While the entanglement degradation with the Gibbons-Hawking states thus seems to be intuitively well acceptable, we wish to present below another viable description of the SdS thermodynamics where such intuitions seems to fail.

\section{The total entropy and the effective temperature}\label{s4} 
\noindent
 The Gibbons-Hawking framework discussed in the preceding section allows us to treat the two event horizons of the Schwarzschild-de Sitter spacetime separately, and provides a thermal description of them in terms of their individual temperatures $\kappa_H/2\pi$ and $\kappa_C/2\pi$. As we mentioned earlier, these individual temperatures correspond to individual entropies $A_H/4$ and $A_C/4$ respectively, Eq.~({l4}). Since the entropy is a measure of lack of information to an observer, for an observer located in region C in Fig.~(\ref{fl1}), one can also define a total entropy $S= (A_{H}+A_C)/4$. Thus for a fixed $\Lambda$,
$$\delta S= \frac14 \left(\frac{\delta A_H}{\delta M}+\frac{\delta A_C}{\delta M} \right)\delta M  $$
Using Eqs.~(\ref{l2}), (\ref{l3}), one then obtains after a little algebra
 a thermodynamic relationship with an  {\it effective equilibrium} temperature~\cite{Maeda, Davies, Saida1, Saida2, Urano, Bhattacharya:2015mja, Zhang, Kanti1, Kanti2, Kanti3, pappas},
\be
\delta M =- \frac{\kappa_H\kappa_C}{2\pi(\kappa_H-\kappa_C)}\delta S=-T_{\rm eff}\delta S
\label{eff}
\ee
 Even though  various computations including that of phase transition has been done using $T_{\rm eff}$, e.g.~\cite{Zhang, Kanti1, Kanti2, Kanti3, pappas}, a clear   understanding of it in terms of  field quantisation and  explicit vacuum states seems to be missing.  

 Let us first try to understand the emergence of this effective temperature intuitively. We note that in this picture where the two horizons are combined, we must consider emission as well as absorption of Hawking radiations, both of which change the horizon areas. For example ignoring the greybody effects, a particle emitted from the BEH will propagate towards CEH and will eventually get absorbed. Likewise a particle emitted from the CEH will propagate inward and will be absorbed by the BEH.  Since $\kappa_H >\kappa_C$, Eq.~(\ref{l3}), the flux of outgoing particles emitted from the BEH at any point $r_H < r<r_C$ will be greater than the flux of particles propagating inward emitted from the CEH, resulting in an effective outward flux and evaporation of the black hole. The existence of such effective temperature can then be intuitively understood as follows. Let {\small $P_E^H(P_A^H$)} and {\small $P_E^C(P_A^C)$} respectively be the single particle emission (absorption) probabilities for the BEH and CEH, so that
 {\small $P^{H}_{E}=P^{H}_{A}e^{-2\pi \omega/\kappa_H}$}  and {\small $P^{C}_{E}=P^{C}_{A}e^{-2\pi \omega/\kappa_C }$}.
By treating the above probabilities as independent, we may define an {\it effective emission probability}, {\small $P^E_{\rm eff}: = P^{C}_{E}P^{H}_{A}= P^{C}_{A}P^{H}_{E}e^{-\omega/T_{\rm eff}}$},  corresponding to the effective inward flux of the cosmological Hawking radiation on the BEH. Likewise the effective absorption probability {\small $P^A_{\rm eff}=P^{C}_{A}P^{H}_{E}$}, corresponds to the   effective outward flux of  black hole's Hawking radiation on the CEH.  Since {\small $P^A_{\rm eff} > P^E_{\rm eff}$}, the black hole gets evaporated.

From Eq.~(\ref{l2}), Eq.~(\ref{l3}), we also have
\be
\lim_{3M\sqrt{\Lambda} \to 0} T_{\rm eff} \to \frac{\kappa_C}{2\pi} \approx \frac{1}{2\pi}\sqrt{\frac{\Lambda}{3}}, \qquad \lim_{3M\sqrt{\Lambda} \to 1} T_{\rm eff} \approx \frac{1}{2\pi}\frac{3\sqrt{\Lambda}}{4}
\label{}
\ee
Thus  even though individually the surface gravities $\kappa_H$ and $\kappa_C$ become vanishing in the Nariai limit ($3M\sqrt{\Lambda} \to 1$), $T_{\rm eff}$ is non-vanishing. Moreover, the effective temperature in the Nariai limit is greater than that of when the black hole is extremely hot ($3M\sqrt{\Lambda} \to 0$, with $\Lambda$ fixed). This corresponds to the fact that $T_{\rm eff}$ is related to an emission probability which corresponds to the inward particle flux on the BEH created by CEH, as described above. As the black hole Hawking temperature increases due to decrease in $3M \sqrt{\Lambda}$, the black hole radiates more resulting in larger outward flux on CEH, corresponding to  reduced  effective inward flux or reduced effective temperature. 

However, a thermodynamic relationship such as Eq.~(\ref{eff}) always needs to be proven via the explicit demonstration of particle creation. 
Accordingly, we now wish to explicitly find out  the   quantum states corresponding to the above description. Since we are treating both the horizons together, it is natural to ask, could there be a global vacuum in L$\cup$C$\cup$R in Fig.~(\ref{fl1}), which plays a role here? The answer is no,  for there exists no analytic Feynman propagator that connects both the horizons~\cite{Gibbons}. This implies that (unlike the single horizon cases) one cannot construct any single global mode which is analytic on or across both the horizons and hence in
L$\cup$C$\cup$R. Such non-existence should be attributed to the fact that   there exists no single Kruskal-like coordinates which remove the coordinate singularities of both the horizons. 

 We recall  that  once we relax the idea of Lorentz invariance such as in a curved spacetime, we have the liberty to choose any coordinate system to describe a given phenomenon, as each such coordinate system represents a viable observer. For example in the Schwarzschild spacetime, one chooses different null coordinates to construct various vacuum states, e.g.~\cite{Choudhury:2004ph} and references therein. Thus we shall now introduce a new coordinate system to address the issue of this effective temperature, as follows.

Note that the `surface gravity' $\kappa_U$, Eq.~(\ref{ds5}),  of the unphysical horizon at $r_U=-(r_H+r_C)$   is given by\, $-\partial_r g_{tt} (r_U)/2$. From Eqs.~(\ref{l3}), it is easy to see that
$$\frac{1}{\kappa_U}= \frac{1}{\kappa_C}-\frac{1}{\kappa_H}$$
Using Eq.~(\ref{ds5}), let us now try to remove the `singularity' of the metric at $r=r_U$. Accordingly, we rewrite the $t-r$ part of Eq.~(\ref{l1}) as  
\begin{eqnarray}
ds^2=-\frac{2M}{r}\left\vert1-\frac{r}{r_C}\right\vert^{1+\frac{\kappa_U}{\kappa_C}} \left\vert\frac{r}{r_H}-1\right\vert^{1-\frac{\kappa_U}{\kappa_H}}\, d{\overline u} d {\overline v}
\label{c4}
\end{eqnarray}
where we have defined,
\begin{eqnarray}
{\overline u}=-\frac{1}{\kappa_U}e^{-\kappa_U u},\quad {\overline v}=\frac{1}{\kappa_U}e^{\kappa_U v} \quad 
\label{c5}
\end{eqnarray}
Apparently it might appear that we are analytically extending the spacetime metric  at $r_U$. However due to the singularity at $r=0$, the spacetime cannot be extended to negative radial values.  Note also that  the metric in Eq.~(\ref{c4}) is {\it not} well behaved on or across any of the event horizons. Thus Eq.~(\ref{c4}), Eq.~(\ref{c5}) {\it do not} correspond to the beyond horizon extensions of Fig.~(\ref{fl1}) and hence they represent a coordinate system only in region C, $r_H < r <r_C$, which is our region of interest anyway. By considering incoming and outgoing null geodesics, it is easy to check that,
$$ u_{\rm in}(r_C) \to  -\infty,~u_{\rm in}(r_H) \to  \infty,\quad v_{\rm out}(r_H) \to -\infty,~v_{\rm out}(r_C) \to \infty $$
yielding $-\infty < \overline{u} \leq 0$ and $0 \leq  \overline{v} < \infty$. It is also easy to see the hyperbolic locus of the $r={\rm const.}$ curves with respect to $\overline{u}$ and $\overline{v}$, as of the previous cases, Eq.~(\ref{ds5'}).

We now define a field quantisation in terms of  $(\overline{u},\overline{v})$  and an alternative one in terms of the usual  $(u,v)$ coordinates as earlier. Using the ranges of coordinates and following the standard procedure, e.g.~\cite{JHT}, we can compute the Bogoliubov coefficients  by choosing the integration surface infinitesimally close to, eg. the BEH. Denoting the vacuum defined by the  $(\overline{u},\overline{v})$ modes by $| \overline{0}\rangle$, we have the squeezed state relationship between the two kind of states corresponding to the above two field quantisations, 
\begin{eqnarray}
| \overline 0\rangle =\sum_{n=0}^{\infty} \frac{\tanh^n {w}}{\cosh{w}}|n, n\rangle,\quad {\rm with }~~\tanh w = e^{-\pi \omega/\kappa_U}
\label{c6}
\end{eqnarray}
the above corresponds to a pair creation with temperature $T_{\rm eff}=\kappa_U/2\pi$.  We emphasise once again that the (entangled) pair creation is occurring in this case only in $r_H < r < r_C$ and {\it not} in the causally disconnected wedges as of the preceding Section. Accordingly, $|\overline{0}\rangle$ should not be regarded as any analogue of the global or Minkowski vacuum.  We also emphasise that the appearance of $\kappa_U$ (instead of $\kappa_H$ or $\kappa_C$) in Eq.~(\ref{c5}) has guaranteed the emergence of the temperature $T_{\rm eff.}$. Due to this reason, the coordinate system in Eqs.~(\ref{c4}),~(\ref{c5}) seems to be unique, as far as this effective description is concerned. Since we are quantising the field using two different coordinatisations ($u,v$ and $\overline{u}, \overline{v}$), the associated vacua are energetically different, giving rise to the  Bogoliubov relationship and particle creation.  Since the global and the local vacua are confined in the single region $r_H < r< r_C$, the particle creation here qualitatively rather resembles with that of in the cosmological spacetimes, e.g.~\cite{Bhattacharya:2020sjr}. 

\begin{figure}[h!]
\begin{center}
  \includegraphics[width=6.5cm]{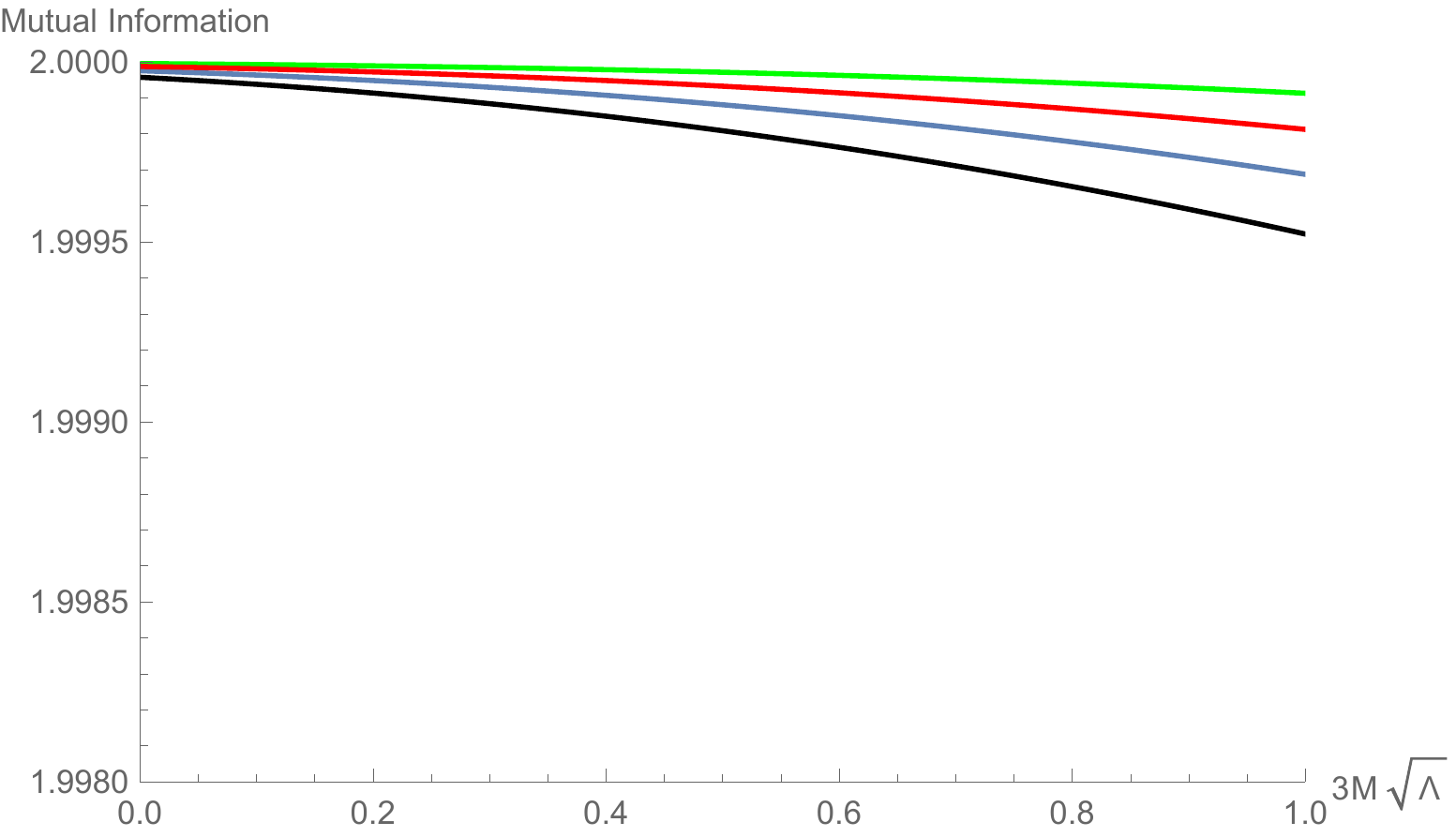}
  \includegraphics[width=6.5cm]{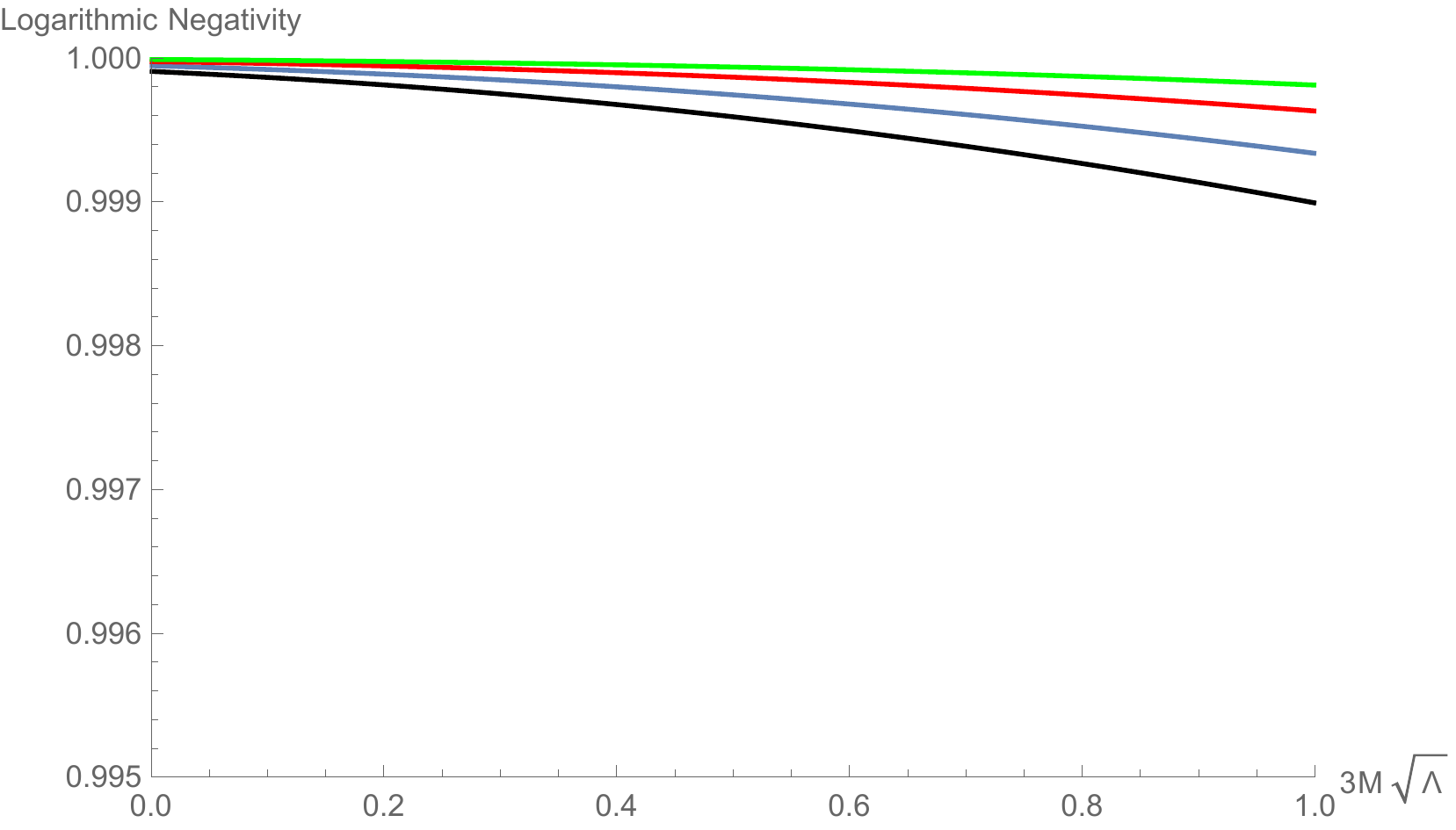}
  \caption{(Left) The mutual information corresponding to the maximally entangled state $|\chi\rangle$, Eq.~(\ref{c7}), vs. $3M\sqrt{\Lambda}$.  (Right) The logarithmic negativity vs. $3M\sqrt{\Lambda}$. The black,  blue, red, green curves correspond respectively to $\omega/ \sqrt{\Lambda}  = 0.95, 1,\,1.07,\,1.15$. Note the complete qualitatively opposite behaviour with respect to the Gibbons-Hawking states, Figs.~(\ref{fig2}). See main text for detail.}

  \label{fig3}
\end{center}
\end{figure}

We now take a maximally entangled state analogous to Eq.~(\ref{x1}),
\begin{eqnarray}
|\chi \rangle = \frac{1}{\sqrt{2}}\left[|\overline{0}, \overline{0}\rangle+|\overline{1},  \overline{1}\rangle\right],
\label{c7}
\end{eqnarray}
and  expand it using Eq.~(\ref{c6}). We then trace out parts of it in order  to form a mixed bipartite system and compute as earlier the mutual information and the logarithmic negativity, plotted in Figs.~(\ref{fig3}). As expected from the characteristics of $T_{\rm eff}$   discussed above, the entanglement does not degrade even when the black hole is extremely hot $(3M \sqrt{\Lambda} \to 0$, for a fixed $\Lambda$), but actually it is maximum in this limit. Moreover, the entanglement degrades as we approach the Nariai limit.  This is completely contrary to what was obtained with the Gibbons-Hawking states in Sec.~3, or to the best of our knowledge, what has been reported in the literature so far, \cite{mann}-\cite{Mirza}. Note also from the figure that (for a fixed $\omega$) the entanglement degrades with increasing $\Lambda$ value, corresponds to the increasing temperature and particle creation by the CEH. In the next Section we have explained that $T_{\rm eff.}$ can have {\it no} analogue in single horizon (i.e. $\Lambda \leq 0$) spacetimes.  
\section{Discussion}
\noindent
We have analysed the entanglement degradation for maximally entanglement Kruskal-like states in the Schwarzschild-de Sitter spacetime, by exploring  two viable descriptions of thermodynamics and particle creation in this background. In Sec.~3, we have taken the Gibbons-Hawking proposition~\cite{Gibbons}, where an observer can be in thermal equilibrium with either of the horizons, by the means of placing a thermally opaque membrane in between the two horizons. We have shown that the entanglement degrades in this case with increasing Hawking temperature of either of the horizons. This is qualitatively similar to the earlier results found for single horizon spacetimes~\cite{mann}-\cite{Mirza}. In Sec.~4, we have addressed the total entropy-effective temperature formalism~\cite{Maeda, Davies, Urano}, in order to treat both the horizons in an equal footing. By introducing a suitable coordinate system, we have found the vacuum states necessary for such description and have shown that  entanglement degradation never happens in this scenario, no matter how hot the black hole becomes or how small the cosmological constant is. Note from  Eq.~(\ref{l2}) that the total entropy is minimum in the Nariai limit ($3M\sqrt{\Lambda} \to 1$) and maximum as $3M\sqrt{\Lambda} \to 0$. Thus if we begin from the Nariai limit (where the Hawking temperature of both horizons are vanishingly small), assuming $\Lambda$ is fixed the black hole will evaporate and the spacetime will evolve towards $3M\sqrt{\Lambda} \to 0$. In this course, we shall keep recovering the entanglement or correlation, Figs.~(\ref{fig3}). This is in complete qualitative contrast with the existing cases~\cite{mann}-\cite{Mirza}.  Finally,   note from Eqs.~(\ref{l3}), Eq.~(\ref{eff}) that $T_{\rm eff} \to 0$ as $\Lambda \to0$. Also, there can be no cosmological event horizon for $\Lambda \leq 0$. This means that $T_{\rm eff}$ and the associated entanglement phenomenon  can have {\it no} $\Lambda \leq 0$ analogue.   Since a black hole spacetime endowed with a positive  $\Lambda$  furnishes a nice toy model for the global structure of a black hole spacetime in the current as well as in the early inflationary universes, these results seem to be physically interesting in its own right.

 Note that the `entanglement degradation'  here (and also in the earlier relevant works~\cite{mann}-\cite{Mirza}) does not correspond to some dynamical decoherence procedure, but rather to some observers who use different coordinate systems and hence different physically well motivated  vacuum states. On the other hand, the decoherence mechanism is likely to involve a more realistic collapse scenario, in order to see the time evolution of a given initial vacuum state. Nevertheless,  since particle creation in an eternal black hole spacetime may effectively model that of in a collapsing geometry~\cite{BD}, one may expect that the above demonstrations indeed have connections to the Hawking radiation and entanglement degradation/decoherence in a  collapsing geometry. However, we are unaware of any such explicit computations and certainly this warrants further attention.

 Although we have simply worked in $1+1$-dimensional Schwarzschild-de Sitter spacetime, the qualitative features of entanglement we have found here would certainly remain the same in higher dimensions.  This is because the Bogoliubov coefficients and hence the particle creation does not depend  upon the angular eigenvalues for a spherically symmetric spacetime, e.g.~\cite{BD}.  Similar analysis for fermions may be interesting. The rotating black hole spacetimes, due to  the existence of various exotic vacuum states for a massless scalar~\cite{Kay:1988mu} also seems to be interesting in this context.  However, perhaps it may be more interesting in this context to consider the acoustic analogue gravity phenomenon~\cite{Barcelo:2005fc}, where the propagation of the perturbation in a fluid is associated with an internal acoustic geometry endowed with a sonic causal structure and horizon. Accordingly, there is creation of phonons with Hawking like spectra. For a multi-component fluid, perhaps  one may naively then  expect a multi-sonic horizon structure analogous to the SdS. For such an acoustic analogue  system, we may look for analogous entanglement properties as that of the the SdS. These construction, like the other analogue gravity phenomenon, might make some interesting predictions testable in the laboratory.  We hope to return to these issues in future works.\\

\noindent
\textbf{Acknowledgements:} SB's research was partially supported by the ISIRD grant 9-289/2017/IITRPR/704. NJ would like to thank H.~Hoshino for his help in the numerical analyses part. We dedicate this manuscript to late Prof. Thanu Padmanabhan.



\begin{thebibliography}{99} 

\bibitem{mann}
I.~Fuentes-Schuller and R.~B.~Mann, {\it Alice falls into a black hole: Entanglement in non-inertial frames},
Phys.~Rev.~Lett.{\bf95} (2005) 120404,
arXiv:quant-ph/0410172




\bibitem{Pan:2008yi}
Q.~Pan and J.~Jing,
{\it Degradation of non-maximal entanglement of scalar and Dirac fields in non-inertial frames},
Phys.~Rev.~A{\bf77}, 024302 (2008)
[arXiv:0802.1238[quant-ph]]

\bibitem{martinez}
E.~Martin-Martinez, L.~J.~Garay, J.~Leon, {\it Unveiling quantum entanglement degradation near a Schwarzschild black hole},
Phys.~Rev.~D{\bf82}, 064006 (2010)
arXiv:1006.1394[quant-ph]



\bibitem{martin}
E.~Martin-Martinez, L.~J.~Garay, J.~Leon, {\it Quantum entanglement produced in the formation of a black hole},
Phys.~Rev.~D{\bf82}, 064028 (2010)
arXiv:1007.2858[quant-ph]

\bibitem{Montero:2011sx}
M.~Montero, J.~Leon and E.~Martin-Martinez,
{\it Fermionic entanglement extinction in non-inertial frames},
Phys. Rev. A{\bf84}, 042320 (2011)


\bibitem{richter}
B.~Richter and Y.~Omar, {\it Degradation of entanglement between two accelerated parties: Bell states under the Unruh effect},
Phys.~Rev.~A{\bf92}, 022334 (2015)
arXiv:1503.07526[quant-ph]

\bibitem{Asghari:2018} 
M.~ Asghari, P.~ Pedram,  and P.~ Espoukeh, 
{\it Entanglement degradation in the presence of the Kerr-Newman black hole},
Quantum~inf.~process{\bf17}, 115 (2018) 

\bibitem{Fuentes:2010dt}
I.~Fuentes, R.~B.~Mann, E.~Martin-Martinez and S.~Moradi,
{\it Entanglement of Dirac fields in an expanding spacetime},
Phys. Rev. D{\bf82}, 045030 (2010)
[arXiv:1007.1569[quant-ph]]

\bibitem{Mirza}
H~Mehri-Dehnavi, R.~R~Darabad, H.~Mohammadzadeh, Z.~Ebadi and B.~Mirza, {\it Quantum teleportation with nonclassical correlated states in noninertial frames},
Quantum Information Processing{\bf14} (3), 1025 (2015)


\bibitem{Bhattacharya:2019zno}
S.~Bhattacharya, S.~Chakrabortty and S.~Goyal,
{\it Dirac fermion, cosmological event horizons and quantum entanglement},
Phys.~Rev.~D{\bf101}, no.8, 085016 (2020)
[arXiv:1912.12272 [hep-th]]


\bibitem{Bhattacharya:2020sjr}
S.~Bhattacharya, S.~Chakrabortty, H.~Hoshino and S.~Kaushal,
{\it Background magnetic field and quantum correlations in the Schwinger effect},
Phys.~Lett.~B{\bf811}, 135875 (2020)
[arXiv:2005.12866[hep-th]]

 
 
  
 
 \bibitem{Gibbons} 
G.~W.~Gibbons and S.~W.~Hawking, {\it Cosmological event horizons, thermodynamics, and particle creation},
Phys.~Rev.~D{\bf15}, 2738 (1977)


 \bibitem{Kastor}
 D.~Kastor and J.~H.~Traschen, {\it Particle production and positive energy theorems for charged black holes in De Sitter}, Class.~Quant.~Grav.{\bf13}, 2753 (1996) [gr-qc/9311025]
 
 \bibitem{Bousso}
R.~Bousso and S.~W.~Hawking, {\it (Anti)evaporation of Schwarzschild-de Sitter black holes}, Phys.~Rev.~D{\bf57}, 2436 (1998) [hep-th/9709224]

\bibitem{Bousso2}
R.~Bousso, {\it Quantum global structure of de Sitter space}, Phys.~Rev.~D{\bf60}, 063503 (1999) [hepth/9902183]



\bibitem{JHT}
J.~H.~Traschen, {\it An Introduction to black hole evaporation}, [arXiv:gr-qc/0010055]

\bibitem{Choudhury:2004ph}
T.~R.~Choudhury and T.~Padmanabhan,
{\it Concept of temperature in multi-horizon spacetimes: Analysis of Schwarzschild-de Sitter metric},
Gen.~Rel.~Grav.\textbf{39}, 1789 (2007)
[arXiv:gr-qc/0404091[gr-qc]].

\bibitem{bhatta}
S.~Bhattacharya, A.~Lahiri, {\it Mass function and particle creation in Schwarzschild-de Sitter spacetime},
Eur.~Phys.~J.~C{\bf73}, 2673 (2013)
[arXiv:1301.4532[gr-qc]]

\bibitem{Bhattacharya:2018ltm}
S.~Bhattacharya,
{\it Particle creation by de Sitter black holes revisited},
Phys.~Rev.~D{\bf98}, no.12, 125013 (2018)
[arXiv:1810.13260[gr-qc]]

\bibitem{qiu}
Y.~Qiu, J.~Traschen, {\it Black Hole and Cosmological Particle Production in Schwarzschild de Sitter},
Class.~Quant.~Grav.{\bf37}, no.13, 135012 (2020)
[arXiv:1908.02737[hep-th]]

 \bibitem{Goheer}
 N.~Goheer, M.~Kleban and L.~Susskind, {\it The Trouble with de Sitter Space}, JHEP 0307, 056 (2003) [arXiv:hep-th/0212209]
 
  \bibitem{Park}
M.~I.~Park, {\it Statistical Mechanics of Three-dimensional Kerr-de Sitter Space
}, Class.~Quant.~Grav.{\bf26}, 075023 (2009) [arXiv:hep-th/0306152]

\bibitem{Dolan1}
B.~P.~Dolan, D.~Kastor, D.~Kubiznak, R.~B.~Mann and J.~Traschen,{\it Thermodynamic Volumes
and Isoperimetric Inequalities for de Sitter Black Holes}, Phys.~Rev.~D{\bf87}, 104017 (2013) [arXiv:1301.5926 [hep-th]]

\bibitem{Dolan2}
B.~P.~Dolan, {\it Black holes and Boyle's law The thermodynamics of the cosmological constant}, Mod.~Phys.~Lett.~A{\bf30}, no. 03n04, 1540002 (2015) [arXiv:1408.4023 [gr-qc]]


%

\bibitem{Maeda}
K.~Maeda, T.~Koike, M.~Narita and A.~Ishibashi, {\it Upper bound for entropy in asymptotically de Sitter
space-time}, Phys.~Rev.~D{\bf57}, 3503 (1998) [gr-qc/9712029]

\bibitem{Davies}
P.~C.~W.~Davies and T.~M.~Davis, {\it How far can the generalized second law be generalized}, Found.~Phys.{\bf32}, 1877 (2002) [astro-ph/0310522]

\bibitem{Urano}
M.~Urano, A.~Tomimatsu and H.~Saida, {\it Mechanical First Law of Black Hole Spacetimes with Cosmological Constant and Its Application to Schwarzschild-de Sitter Spacetime}, Class.~Quant.~Grav.{\bf26}, 105010 (2009)  [arXiv:0903.4230[gr-qc]]


\bibitem{Saida1}
H.~Saida, {\it de Sitter thermodynamics in the canonical ensemble}, Prog.~Theor.~Phys.{\bf122},  1239 (2010) [arXiv:0908.3041[gr-qc]]

 \bibitem{Saida2}
H.~Saida, {\it To what extent is the entropy-area law universal? : Multi-horizon and multi-temperature
spacetime may break the entropy-area law}, Prog.~Theor.~Phys.{\bf122}, 1515 (2010)  [arXiv:0910.2510[gr-qc]]

 
\bibitem{Bhattacharya:2015mja}
S.~Bhattacharya,
{\it A note on entropy of de Sitter black holes},
Eur. Phys. J. C{\bf76}, no.3, 112 (2016)
[arXiv:1506.07809[gr-qc]]

\bibitem{Zhang}
L.~C.~Zhang, M.~S.~Ma, H.~H.~Zhao and R.~Zhao, {\it Thermodynamics of phase transition in higherdimensional Reissner-Nordstr\"{o}m-de Sitter black hole}, Eur.~Phys.~J.~C{\bf74}, no. 9, 3052 (2014)
[arXiv:1403.2151[gr-qc]]

\bibitem{Kanti1}
P.~Kanti, T.~Pappas and N.~Pappas, {\it Greybody factors for scalar fields emitted by a higher-dimensional
Schwarzschild-de Sitter black hole}, Phys.~Rev.~D{\bf90}, no. 12, 124077 (2014) [arXiv:1409.8664[hep-th]]

\bibitem{Kanti2}
T.~Pappas, P.~Kanti and N.~Pappas, {\it Hawking radiation spectra for scalar fields by a higher-dimensional
Schwarzschild-de Sitter black hole}, Phys.~Rev.~D{\bf94}, no. 2, 024035 (2016) [arXiv:1604.08617[hep-th]]

\bibitem{Kanti3}
P.~Kanti and T.~Pappas, {\it Effective temperatures and radiation spectra for a higher-dimensional Schwarzschild-de Sitter black hole}, Phys.~Rev.~D{\bf96}, no. 2, 024038 (2017) [arXiv:1705.09108[hepth]]

\bibitem{pappas}
T.~Pappas, P.~Kanti, {\it Schwarzschild-de Sitter spacetime: the role of Temperature in the emission of Hawking radiation},
Phys.~Lett.~B{\bf775}, 140 (2017)
arXiv:1707.04900[hep-th]


\bibitem{BD}
N.~D.~Birrell and P.~C.~W.~Davies,
{\it Quantum Fields in Curved Space}, Cambridge Univ. Press (UK) (1984).


  
  \bibitem{Unruh:1976db}
W.~G.~Unruh,
{\it Notes on black hole evaporation},
Phys.~Rev.~D{\bf14}, 870 (1976)

\bibitem{Crispino:2007eb}
L.~C.~B.~Crispino, A.~Higuchi and G.~E.~A.~Matsas,
{\it The Unruh effect and its applications},
Rev.~Mod.~Phys.{\bf80}, 787 (2008)
[arXiv:0710.5373 [gr-qc]]
  
  \bibitem{NC} 
 M.~A.~Nielsen and I.~L.~Chuang, {\it Quantum Computation and Information Theory}, Cambridge Univ. Pr., UK (2010)


\bibitem{Kay:1988mu}
B.~S.~Kay and R.~M.~Wald,
{\it Theorems on the Uniqueness and Thermal Properties of Stationary, Nonsingular, Quasifree States on Space-Times with a Bifurcate Killing Horizon},
Phys.~Rept.{\bf 207}, 49 (1991)


\bibitem{Barcelo:2005fc}
C.~Barcelo, S.~Liberati and M.~Visser,
{\it Analogue gravity},
Living Rev.~Rel.\textbf{8}, 12 (2005)
[arXiv:gr-qc/0505065 [gr-qc]].

  
\end{thebibliography}
\end{document}